# Complex surface spin polarization of the $La_{0.7}Sr_{0.3}MnO_3$ films

M. Cavallini, F. Biscarini, P. Nozar, V. Dediu[@]

*Consiglio Nazionale delle Ricerche, Istituto per lo Studio dei Materiali Nanostrutturati,*
*via P. Gobetti 101, I-40129 Bologna, Italy*

## ABSTRACT

The surface spin polarization in the epitaxial films of a prototype manganite, $La_{0.7}Sr_{0.3}MnO_3$, has been investigated by Scanning Tunneling Spectroscopy with ferromagnetic SP tips at room temperature. The manganite surface splits into ferromagnetic (FM) and paramagnetic (PM) regions characterized by metallic and insulating like behavior respectively. Spin polarized spectroscopy with Ni tips has been performed separately for the two phases, and the results compared with standard W and Pt tips. While PM exhibits featureless tunneling characteristics, the FM regions exhibit at room temperature strongly nonlinear behavior with a band-gap like behavior. The spin resolved density of states of the FM regions of the manganite has been extracted by the deconvolution of the spectroscopic curves. It indicates very high spin polarization (nearly half-metallic behavior) with the spin down band separated from Fermi level by 0.4 eV. This value corresponds exactly to the gap value measured for the whole surface of this manganite at low temperatures.

Thus the surface of the $La_{0.7}Sr_{0.3}MnO_3$ films maintain even at room temperature nano- and micrometric islands of very high spin polarization, while the total polarization is given by the coverage of the surface by FM regions. This result is extremely important for spintronic applications of this material and indicates possible roots for the realization of manganite nanosized devices with extremely high spin polarization.

Since the discovery of the colossal magnetoresistance effects (CMR)[1], the manganites $La_{1-x}(Sr, Ca)_xMnO_3$ were expected to play an important role in spintronic devices. The La-Sr based manganite was the first material whose half-metallic properties were demonstrated in a straightforward way by spin resolved photoemission spectroscopy[2] at low temperatures (40 K). The measurements of the spin polarization in such complicated oxides are strongly prevented by the difficulties to prepare clean surfaces. Increasing the films quality, proves for strong spin polarization at higher temperature have appeared in the last years[3-5] mainly from the magnetic tunnel junctions investigations. Among most important results in this direction was the straightforward demonstration of the room temperature spin polarization of LSMO surface in spin tunneling devices[6] and hybrid organic/inorganic spin injection devices[7, 8]. Moreover, in the field of organic spintronics the manganites seem to be by far the most efficient spin injectors[7-9].

The knowledge of distribution of the spin polarization on the LSMO surface becomes thus of crucial importance. The question is also driven by the STM observation of the phase separation at the manganite film surfaces in metallic ferromagnetic parts and insulating paramagnetic ones[10, 11]. Such a distribution should unequivocally induce inhomogeneous spin polarization at the surface.

In this work we present the investigation of magnetic homogeneity and spin polarized properties of the epitaxial $La_{0.7}Sr_{0.3}MnO_3$ film surfaces at room temperature by SP Scanning Tunneling Microscopy. We confirm the surface phase separation[10, 11] in ferromagnetic and paramagnetic regions. Assuming that the temperature dynamics of this phase separation observed by Becker et al.[10, 11] is a common feature of the manganite film surfaces, the investigation of the spin polarization at any given temperature (room T in our case), together with mentioned dynamics, should provide the full set of knowledge for the description of the spin surface properties at any $T<T_C$.

Epitaxial $La_{0.7}Sr_{0.3}MnO_3$ 100 nm thick films have been prepared by pulsed electron ablation technique on $NdGaO_3$ substrates[12]. The films exhibit high $T_c$ (up to 350 K from SQUID and MOKE characterizations) and resistivity lower than 10 mΩcm at 300 K. Detailed surface characterizations were performed by Photoemission spectroscopy (PES) and X-ray Absorption Spectroscopy data at the Mn $L_{2,3}$ edge (XAS). All the investigated films indicated the presence of a surface metallic phase and routinely detectable strong magnetic dichroism at room temperature[13]. All the reported above characterizations have in common a "macroscopic" approach: even PES and XAS methods that are strongly constrained at the surface (within about 1 nm), have a spot size of the order of tens of microns. This holds also for manganite investigation by means of tunneling junctions. Thus the electronic and magnetic properties of the manganite films at the nano-scale remain unclear.

The use of Scanning Tunneling Microscopy (STM) and Spectroscopy (STS) provides a unique opportunity to characterize the surface at nano-scale level, while the use of spin polarized (SP) tips adds a magnetic degree of freedom to this characterization.

In SP STS the maximum transmissivity of carriers between two spin polarized electrodes (tip and sample) separated by a tunnel junction occurs for their parallel spin polarization and drops with increasing misalignment (spin-valve effect)[14-16]. In our experiments STS was operated in parallel tip-sample plane magnetic configuration. For each bias voltage $V_{bias}$, we acquired both topography z(x,y) and *dI(x,y)/dV* (differential tunneling conductance) maps by applying a low-frequency modulation on the $V_{bias}$ while holding the feedback. The simultaneous acquisition of topography and conductivity required a positive bias voltage, limiting our investigations to the tunneling of electrons from Ni tip into the empty states of the manganite.

We used electrochemically etched Ni wire tips[17, 18] as injectors of carriers with strong spin polarization[19, 20], and for comparison Pt and W tips. The reported experiments were performed at room temperature.

The typical morphology of a 100-nm thick film exhibits a smooth background (less than 5 nm roughness on 8 μm lengthscale) with a low density of about 30 nm tall outgrowths (Fig. 1a). No outgrowths are present in the 20 nm thick film (Fig. 1c). The *dI(x,y)/dV* maps (Fig. 1b, d) which are sensitive to local conductance show the coexistence of well defined high conductance (HC) and low conductance (LC) regions. The *dI(x,y)/dV* contrast between LC and HC regions amounts to more than one order of magnitude. In thick films the LC defects are mainly localized on the outgrowths, but not all the outgrowths are LC (Fig. 1b), while the LC are also detected in thin outgrowthless films (Fig. 1d).

Such a phase separation of the manganite film surface has been reported previously[10, 11], and HC and LC have been identified as ferromagnetic and paramagnetic phases correspondingly. Within both the HC and LC regions the conductance is very homogeneous. It is important to mention that this phase separation can be observed by various tips, including non-magnetic ones, with the only difference that the spin polarized tips showed stronger contrast due the magnetic nature of this effect.

The temperature behavior of this phase separation has been investigated in details by Becker et al.[10, 11] that used non-magnetic PtIr tips. It was found that the HC phase dominates at $T<<T_C$, while LC phase dominates near $T_C$. The most interesting fact is that the coverage of the surface by HC phase decreases with temperature in a way very similar to sample magnetization. This indicates a peculiar magnetic transition with strong (low temperature) ferromagnetic order conserved in nano- or microdomains, domains getting smaller and smaller while approaching the Curie temperature. In this fashion the description of the surface spin polarization at any given temperature requires the knowledge of the spin polarization inside the LC and HC phases plus the phases evolution with temperature. If correct, this statement should be confirmed by very high spin polarization of the HC at room temperatures.

The surface spin polarization of the HC and LC regions on $dI(x,y)/dV$ maps have been studied from the evolution of the local conductance versus $V_{bias}$ (Fig. 2). A new method for processing the spectroscopic data was used in order to obtain a higher accuracy with respect to derivatives taken on single pixels. For each $V_{bias}$, a pixel-by-pixel average of $dI(x,y)/dV$ on LC and HC regions is performed. Finally, the $dI(x,y)/dV$ values are normalized to $I/V_{bias}$ to yield an estimate of $d(lnI)/d(lnV)$, which is related to the electronic density of states (DOS). It has been shown that such a procedure removes the effect of voltage and tip-sample separation distance on tunneling current[21].

The *d(lnI)/d(lnV)* curves in Fig. 2 clearly show different responses from the HC and LC regions for the Ni tip. HC characteristics lie above the LC ones at any bias voltage. The LC curve is smooth and extrapolates to zero at $V_{bias} = 0$ indicating an insulating behavior. The HC curve on the other hand, exhibit a metallic and non-linear behavior while its most characteristic feature is a sharp increase at 1.3-1.5 V. The curve is also characterized by some fine structure, like a small narrow peak at 1 V and two local maxima at 1.9 and 2.6 V. A good agreement between data taken with different Ni tips on HC regions of different films was found. No difference was observed between annealed and as-deposited films, apart for the fractional coverage by LC defects.

The comparison with non spin polarized W and Pt tips studies confirmed that the non linearity is due to spin polarized effects: the spectroscopic curves taken by W and Pt tips on the HC regions are featureless and did not indicate any non linear behavior (Fig. 3).

In the next step we perform the manganite bands decovonlution by using available data for the Ni spin dependent DOS. The tip-sample current is calculated by taking into account two separate spin-up and spin-down channels[22]. Thus, the general formula for tunneling current[23] is modified into:

$$I_{\uparrow(\downarrow)} = T \int_0^{eU} N_{\uparrow(\downarrow)}(E)\, N'_{\uparrow(\downarrow)}(E-eU)\, f(E-E_f)\, f(E-E_f-eU)\, dE \qquad (1)$$

Here *T* is the transmission coefficient of the tunnel contact, $N_{\uparrow(\downarrow)}(E)$, $N'_{\uparrow(\downarrow)}(E)$ are spin polarized densities of states of Ni and manganite respectively, and *f(E)* is the Fermi-Dirac distribution function. The most important approximations are to consider *T* independent on the energy *E* and spin orientation. This approximation influences mainly the contrast between spin-up and spin-down channels and cannot lead to any significant shift for the DOS picture. The total current $I_{\uparrow\downarrow}$ has been calculated as the sum of spin-up $I_\uparrow$ and spin-down $I_\downarrow$ currents:

$$I_{\uparrow\downarrow} = I_\uparrow + I_\downarrow \qquad (2)$$

The differential conductivity $d(lnI)/d(lnV)$ was calculated numerically, taking into account the constant current mode experimental procedure.

The spin polarized DOS of Ni is known from both theoretical and experimental investigations. In our deconvolution procedure the SP DOS for bulk[24] and surface[25] Ni states were used (Fig. 4). The use of orientation independent SP DOS for the Ni tip provides a reasonable approach for the case of polycrystalline tips. Even for single crystalline tips the tunneling process will not follow exactly a given crystalline axis direction, due to finite tip curvature and non zero film roughness.

In our approximation the manganite SP DOS was restricted to the 3$d$ bands, namely two-lorentzian ($e_g$ and $t_{2g}$) bands for both spin orientation. The STM method does not distinguish between $e_g$ and $t_{2g}$ symmetries, so we will keep $e_gt_{2g}$ notation for the calculated band structure. Fitting with elliptical and gaussian manganite bands showed no qualitative difference. The choice of lorentzian bands is justified however by the best data-fit agreement and has been successfully used previously for manganite DOS calculations[26].

At positive biases, electrons move from the tip to the manganite, so that only the empty states in the manganite bands are relevant. The Fermi energy position, the band widths, and the separation between the four lorentzians (two spin-up, and two spin-down) were taken as fitting parameters.

The resulting manganite band structure is represented in Fig. 5. It generates the solid (bulk Ni DOS) and dashed (surface Ni DOS) fitting curves for $d(lnI)/d(lnV)$ in Fig. 2. A rather good agreement with experimental data is found for both cases.

For both Ni bulk and surface DOS approximations the fitting is extremely satisfactory. At low biases it provides somehow better agreement for bulk DOS approach, while at hich voltages the agreement with experimental data is better for the surface Ni DOS.

The only sub-band with non-zero DOS at $E_F$ of manganite is the tail of the spin-up polarized band (Fig. 5), which extends roughly up to 1.5 eV (the energies are calculated from $E_F$). This band is associated with the tail of the spin up $e_g t_{2g}$ band of Mn[2, 27]. The spin down band is an overlap of two lorentzian sub-bands, and has non-zero DOS from 0.4 eV up to more than 4 eV. The band width is roughly 2 eV wide, and it is characterized by a maximum at 2.5 eV and a shoulder at 1.8 eV.

A striking feature of the manganite band structure is the presence of the spin gap found for FM regions. The value of this gap, 0.4 eV, coincides exactly with what found by the Fert group[4] in tunnel junctions at low temperature. The meaning of our finding is that the very high spin polarization characteristic for the manganite surface at low temperatures, persists up to high temperatures inside the FM regions, which size are controlled by the M(T) like temperature evolution[10].

Thus, at high temperatures (below $T_C$) there exist channels with nearly half metallic like properties. Taking into account the used approximations we estimate the spin polarization at the Fermi level to be at least higher than 90%: this is the first demonstration of the nearly half metallic properties of manganites at room temperature.

The shape and characteristic energies of derived bands are in good agreement with theoretical ab-initio calculations and optical data[27, 28]. A remarkable agreement between proposed spin down band and $t_{2g}e_g$ spin up d-band re-calculated from spin polarized photoemission should be also noticed[2].

The adopted here approximations (energy independent tunneling coefficient, two lorentzian bands, and orientation averaged DOS for Ni tip) provided the possibility to reveal only the main features of the manganite band structure. A further improvement (band fine substructure) can be obtained on one hand by using single crystalline tips of different spin polarized materials, and, on the other hand, by more sophisticated numerical approach.

In conclusion, the proposed scenario for the spin polarization surface distribution represents a radical modifications of the standard approaches and explains well the high spin injection efficiency of manganites detected in various devices at room tempearture. We believe it offers opportunities for the selection of high spin polarization regions by appropriate nanolithography treatments.

1       S. Jin, H. Tiefel, M. McCormack, et al., Science **264**, 412 (1994).

2       J.-H. Park, E. Vescovo, H.-J. Kim, et al., Nature **392**, 794 (1998).

3       V. Garcia, M. Bibes, A. Barthelemy, et al., Phys. Rev. **B 69**, 052403 (2004).

4       M. Bowen, A. Barthelemy, M. Bibes, et al., Phys. Rev. Lett. **95**, 137203 (2005).

5       M. Bowen, M. Bibes, A. Barthelemy, et al., Appl. Phys. Lett. **82**, 233 (2003).

6       Y. OGIMOTO, M. IZUMI, A. SAWA, et al., Jpn.J.Appl.Phys. **42**, L 369 –L 372 (2003).

7       S. Majumdar, R. Laiho, P. Laukkanen, et al., Appl. Phys. Lett. **89**, 122114 (2006).

8       V. Dediu, M. Murgia, F. C. Matacotta, et al., Solid State Commun. **122**, 181 (2002).

9       Z. H. Xiong, D. Wu, Z. V. Vardeny, et al., Nature **427**, 821 (2004).

10      T. Becker, C. Streng, Y. Luo, et al., Phys. Rev. Lett **89**, 237203 (2002).

11      M. Faeth, S. Freisem, A. A. Menovsky, et al., Science **285**, 1540 (1999).

12      V. Dediu, J. Lòpez, F. C. Matacotta, et al., Phys. Stat. Sol. (b) **215**, 625 (1999).

13      M. P. de Jong, Dediu, V. A., Taliani, C., Salaneck, W. R., J. Appl. Phys. **94**, 7292 (2003).

14      A. Kubetzka, M. Bode, O. Pietzsch, et al., Phys. Rev. Lett. **88**, 057201 (2002).

15      R. Akiyama, H. Tanaka, T. Matsumoto, et al., Appl. Phys. Lett. **79**, 4378 (2001).

16      S. F. Alvarado, Phys. Rev. Lett. **75**, 513 (1995).

17      M. Cavallini and F. Biscarini, Rev. Sci. Instrum. **71**, 4457 (2000).

18      C. Albonetti, I. Bergenti, M. Cavallini, et al., Rev. Sci. Instrum. **73**, 4254 (2002).

19      S. F. Alvarado and P. Renaud, Phys. Rev. Lett. **68**, 1387 (1992).

20      V. P. LaBella, D. W. Bullock, Z. Ding, et al., Science **292**, 1518 (2001).

21      J. A. Stroscio, R. M. Feenstra, and A. P. Fein, Phys. Rev. Lett. **57**, 2579 (1986).

22      J. C. Slonczewski, Phys. Rev. B **39**, 6995 (1989).

23      V. A. Ukraintsev, Phys. Rev. B **53**, 11176 (1996).

24      J. W. D. Connolly, Phys. Rev. **159**, 415 (1967).

25      N. Papanikolaou, cond-mat/0210551 (2002).

[26] D. M. Edwards, A. C. M. Green, and K. Kubo, Physica B **259-261**, 810 (1999).

[27] S. Satpathy, Z. S. Popovic, and F. R. Vukajlovi, Phys. Rev. Lett. **76**, 960 (1996).

[28] A. Chattopadadhyay, A. J. Millis, and S. Das Sarma, Phys. Rev. B **61**, 10738 (2000).

**Figure captions:**

Figure 1

(a) STM topography of a high roughness (80 nm peak-to-peak) zone of the manganite surface and corresponding *d(lnI)/d(lnV)* map (b) acquired simultaneously; (c) and (d) similar images for a low roughness (15 nm peak-to-peak) zone of the manganite surface.

Figure 2

Logarithmic derivatives *d(lnI)/d(lnV)* = *U/I\*dI(x,y)/dV* acquired with Ni tips at *I* = 1 nA and different bias voltages. Both high conductance (HC) and low conductance (LC) regions are shown. The fitting curve represents the results of deconvolution calculations with Ni DOS for bulk (solid line) and surface (dashed line) cases.

Figure 3

Logarithmic derivatives *d(lnI)/d(lnV)* = *U/I\*dI(x,y)/dV* acquired with Pt and W tips at *I* = 1 nA and different bias voltages.

Fig. 4

Spin polarized Ni DOS for bulk (solid line) and surface (dashed line) cases.

Fig. 5

$La_{0.7}Sr_{0.3}MnO_3$ spin polarized DOS at room temperature from the deconvolution of tunneling spectroscopy data in HC regions.

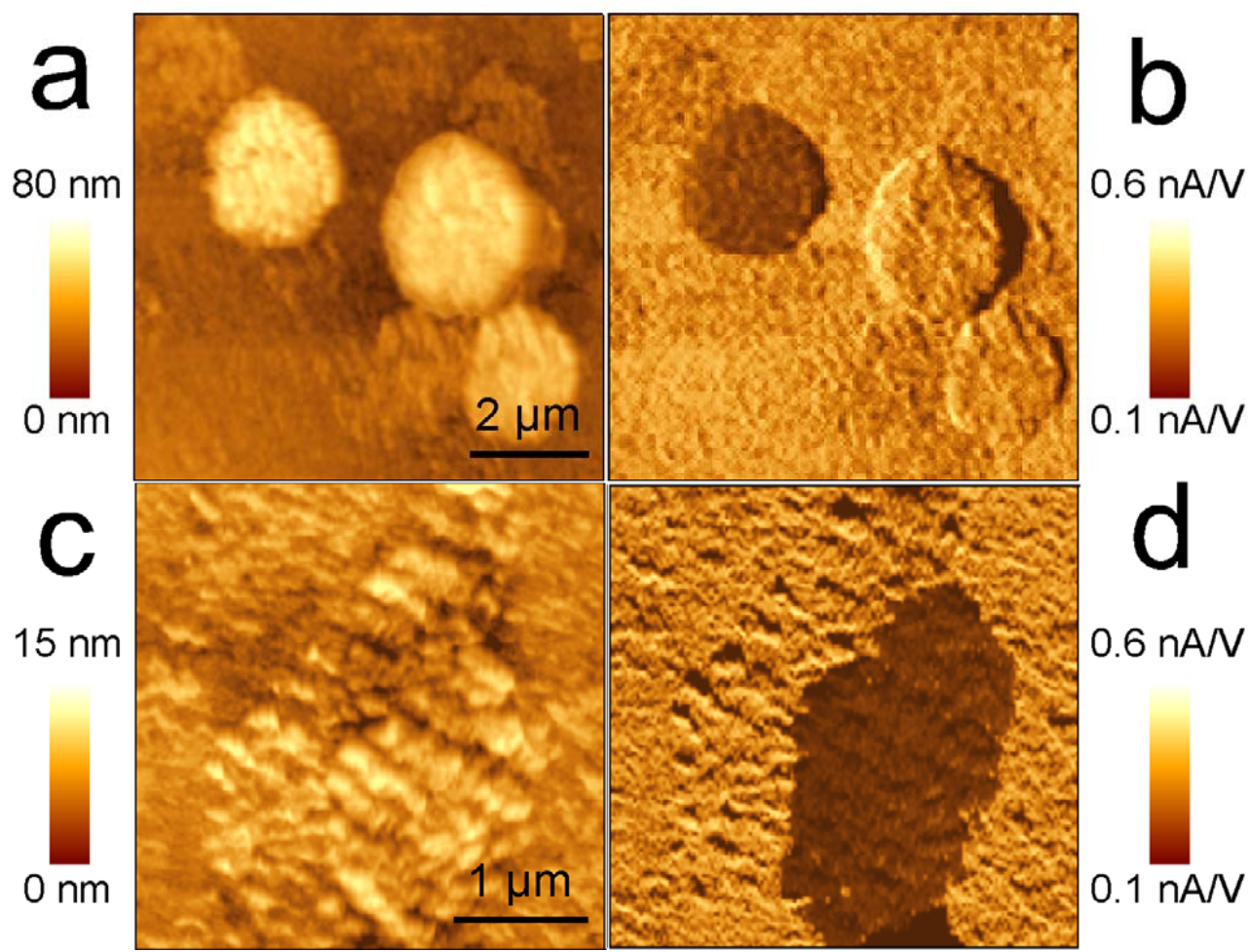

Fig. 1

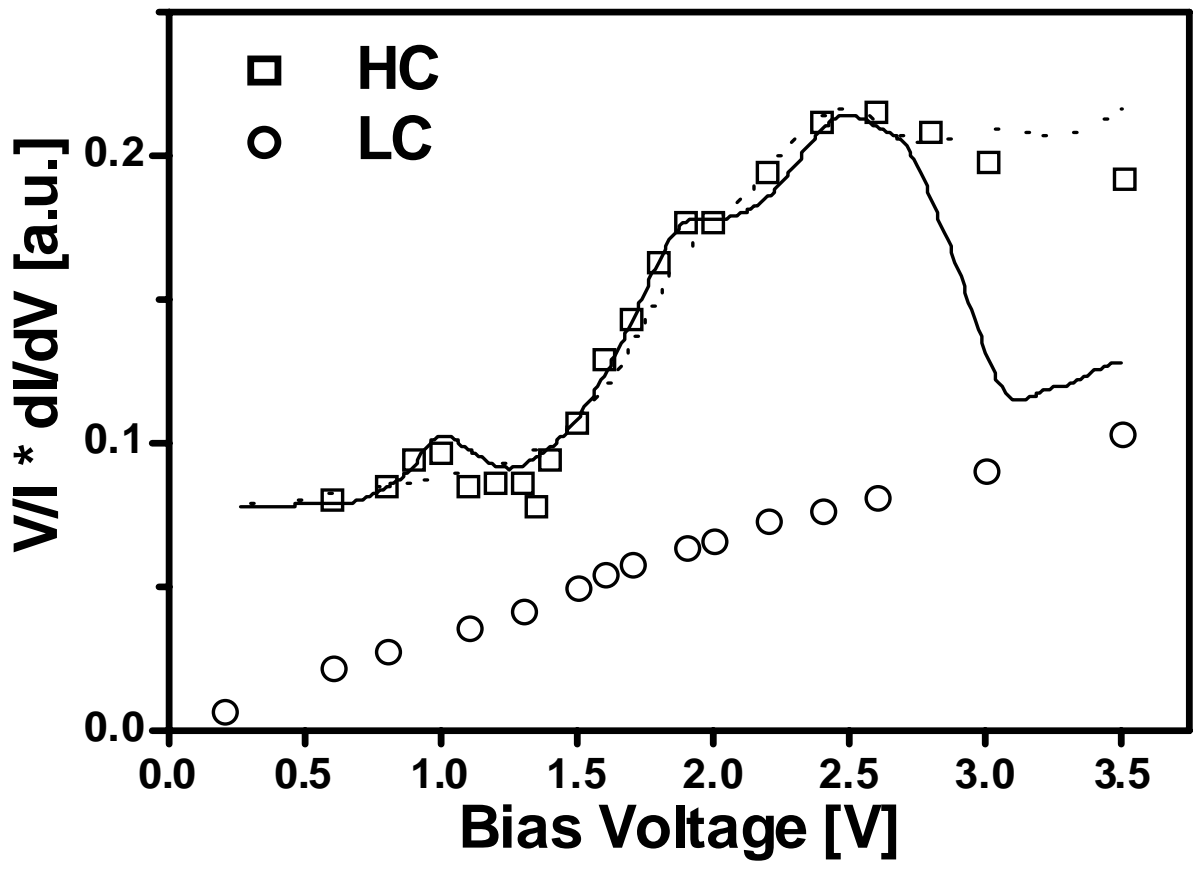

Fig. 2

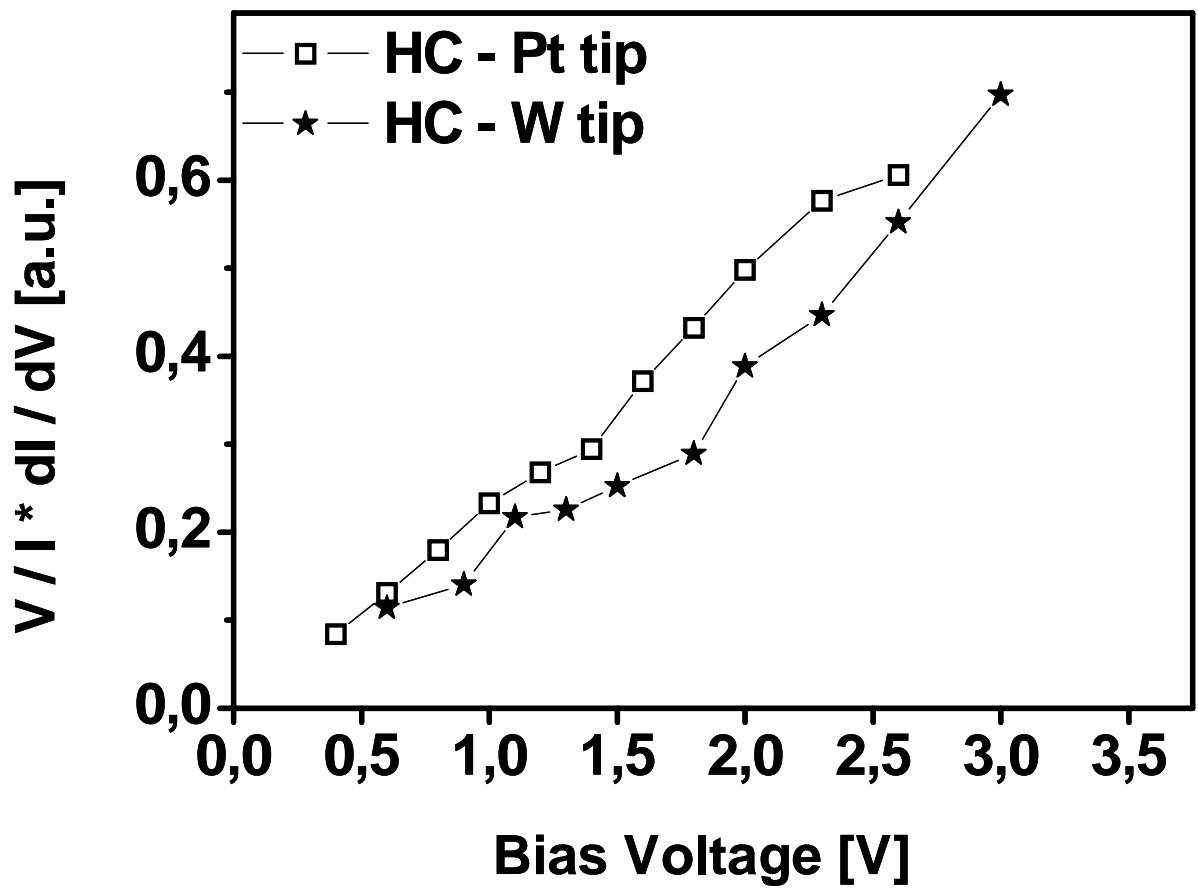

Fig. 3

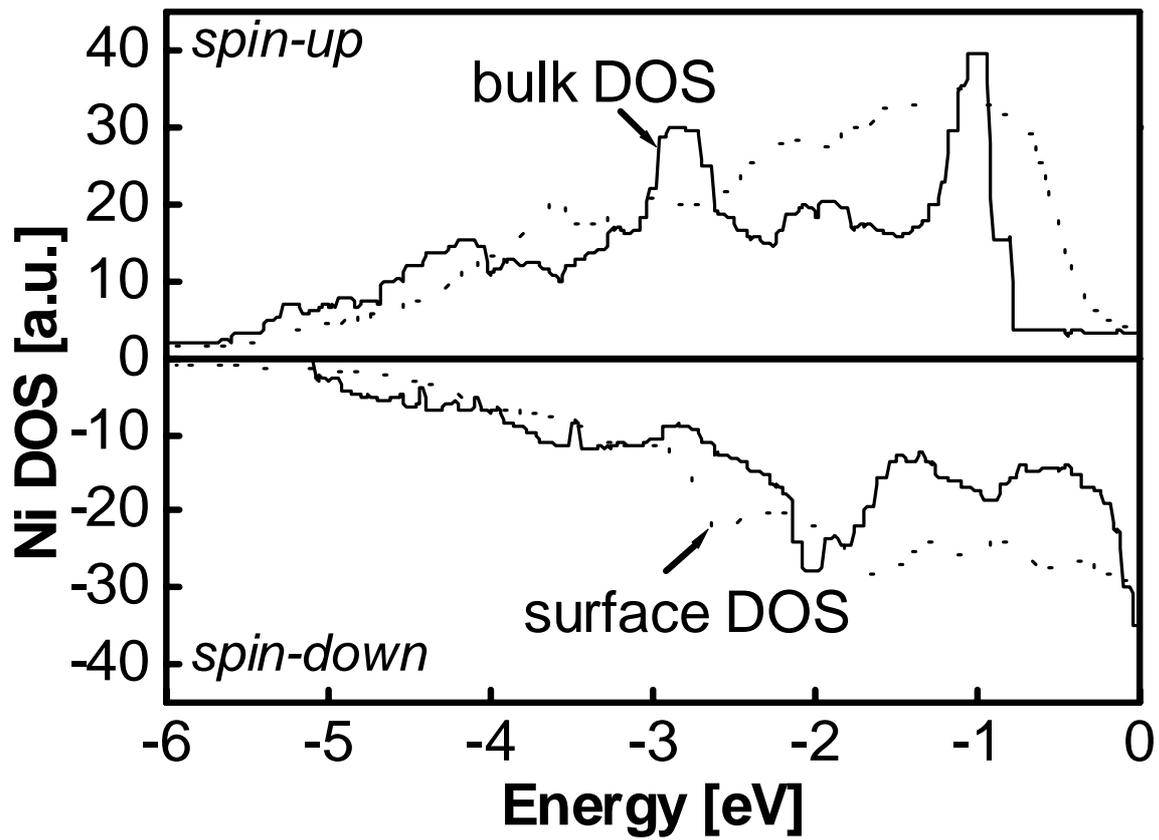

Fig. 4

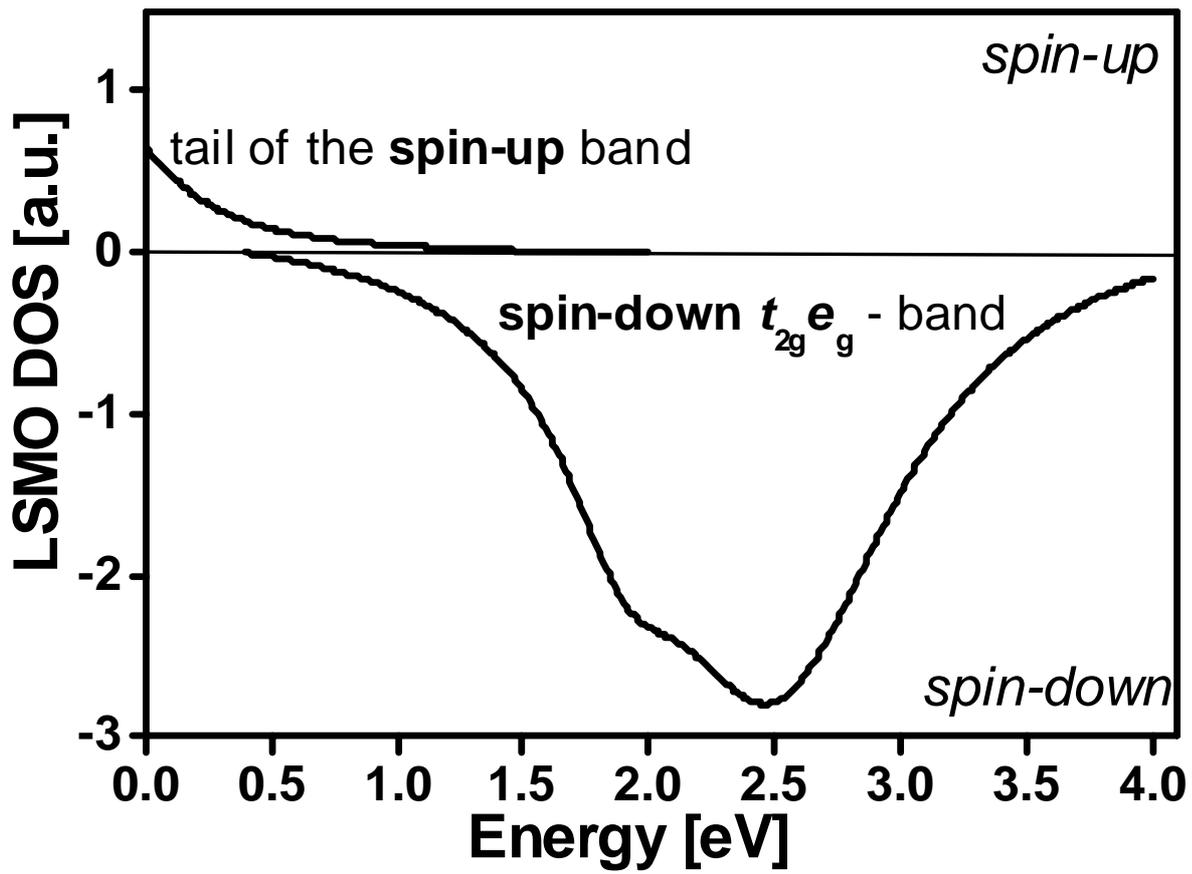

Fig. 5